\documentclass[twocolumn]{article}
\usepackage[a4paper, total={180mm, 243mm}]{geometry}
\usepackage[utf8]{inputenc}
\usepackage{amssymb}
\usepackage{soul}
\usepackage{graphicx}

\usepackage[numbers,sort&compress]{natbib}
\usepackage{siunitx}
\usepackage[version=4]{mhchem} 
\usepackage{titling}
\usepackage{circledsteps}   % This package is for circling text e.g. : \Circled{1}
\usepackage{subcaption}
\usepackage{mathtools}
\usepackage[switch]{lineno} 
\modulolinenumbers[5]

\DeclareUnicodeCharacter{2009}{\,} 
\DeclareCaptionLabelSeparator{custom}{\textbar}
\DeclareCaptionFormat{custom}{\textsf{\textbf{#1 #2} \small #3}}
\newcommand\authormark[1]{\textsuperscript{#1}}

\makeatletter
\newcommand{\showfontsize}{\f@size{} pt}
\makeatother

\definecolor{Myblue}{cmyk}{0.8,0.6,0,0.1}

\title{Femtosecond pulse amplification on a chip}
\author{Mahmoud A. Gaafar\authormark{1,*}, Markus Ludwig\authormark{1,*}, Kai Wang\authormark{2}, Thibault Wildi\authormark{1}, \\
Thibault Voumard \authormark{1}, 
Milan Sinobad\authormark{1}, 
Jan Lorenzen\authormark{1}, 
Henry Francis\authormark{3},
Jose Carreira\authormark{3}, \\
Shuangyou Zhang\authormark{4},
Toby Bi\authormark{4,5}, 
Pascal Del'Haye\authormark{4,5}, 
Michael Geiselmann\authormark{3}, \\
Neetesh Singh\authormark{1},
Franz X. Kärtner\authormark{1,6}, 
Sonia M. Garcia-Blanco\authormark{2},  Tobias Herr\authormark{1,6,**}
 }

\date{%
    \small $^1$Deutsches Elektronen-Synchrotron DESY, Notkestr. 85, 22607 Hamburg, Germany \\
    \small $^2$Integrated Optical Systems, MESA+ Institute for Nanotechnology, University of Twente, 7500AE, Enschede, The Netherlands\\
    \small $^3$LIGENTEC SA, EPFL Innovation Park, Chemin de la Dent-d’Oche 1B, Switzerland CH-1024 Ecublens, Switzerland\\
    \small $^4$ Max-Planck Institute for the Science of Light, 91058 Erlangen, Staudtstr. 2, Germany\\  
    \small $^5$ Department of Physics, FAU Erlangen-Nürnberg, 91058 Erlangen, Germany\\      
    \small $^6$ Department of Physics, Universität Hamburg, Luruper Chaussee 149, 22761 Hamburg, Germany\\  
    \small $^*$These authors contributed equally.\\
    \small $^{**}$tobias.herr@desy.de
}

\begin{document}

\maketitle

% Abstract
\textbf{
    Femtosecond laser pulses enable the synthesis of light across the electromagnetic spectrum and provide access to ultrafast phenomena in physics, biology, and chemistry. Chip-integration of femtosecond technology could revolutionize applications such as point-of-care diagnostics, bio-medical imaging, portable chemical sensing, or autonomous navigation. 
    However, current chip-integrated pulse sources lack the required peak power and on-chip amplification of femtosecond pulses has been an unresolved challenge.
    Here, addressing this challenge, we report $>$50-fold amplification of 1~GHz-repetition-rate chirped femtosecond pulses in a CMOS-compatible photonic chip to 800~W peak power with 116~fs pulse duration. This power level is 2-3 orders of magnitude higher compared to those in previously demonstrated on-chip pulse sources and can provide the power needed to address key applications. To achieve this, detrimental nonlinear effects are mitigated through all-normal dispersion, large mode-area and rare-earth-doped gain waveguides. These results offer a pathway to chip-integrated femtosecond technology with peak power-levels characteristic of table-top sources.
}

%================================
%Introduction and motivation
%================================

\subsection*{Introduction}
Femtosecond laser pulses with high-peak power are a fundamental resource in photonics. By temporally concentrating laser light, they provide unique access to ultrashort time-scales and nonlinear optical effects and have found applications across many disciplines including bio-medical imaging \cite{zipfel_nonlinear_2003, larson_multiphoton_2011, chung_multimodal_2019}, femtochemistry \cite{weiner_femtosecond_1990, zewail_femtochemistry_2000}, optical precision spectroscopy, molecular sensing, low-noise signal generation and time keeping \cite{picque_frequency_2019, fortier_20_2019, diddams_optical_2020}.
Recent developments of photonic-chip integration of femtosecond sources via microresonators \cite{kippenberg2018, pasquazi_micro-combs_2018, gaeta2019}, on-chip mode-locked lasers \cite{van_gasse_recent_2019, chang_integrated_2022,singh_towards_2020,guo_ultrafast_2023}, and integrated electro-optic pulse generators \cite{yu_integrated_2022}, have created tremendous opportunities for femtosecond technology to become accessible at low-cost with high-volume, compact footprint and beyond specialized laboratories. These developments hold potential for large-scale societal benefits such as advanced point-of-care diagnostics, distributed environmental monitoring, and mobile application such as signal generation for navigation and communication.
A current roadblock, however, is the low output peak power of chip-integrated femtosecond sources (around 1~W, or much below that), which is often too low for their envisioned application.
Thus, almost all proof-of-concept demonstrations with ultrashort pulses from on-chip sources have critically relied on table-top optical amplifiers. Integrating these pulse amplifiers on a chip, is a long standing challenge.
\begin{figure*}[t!]\centering\includegraphics[width=0.85\paperwidth]{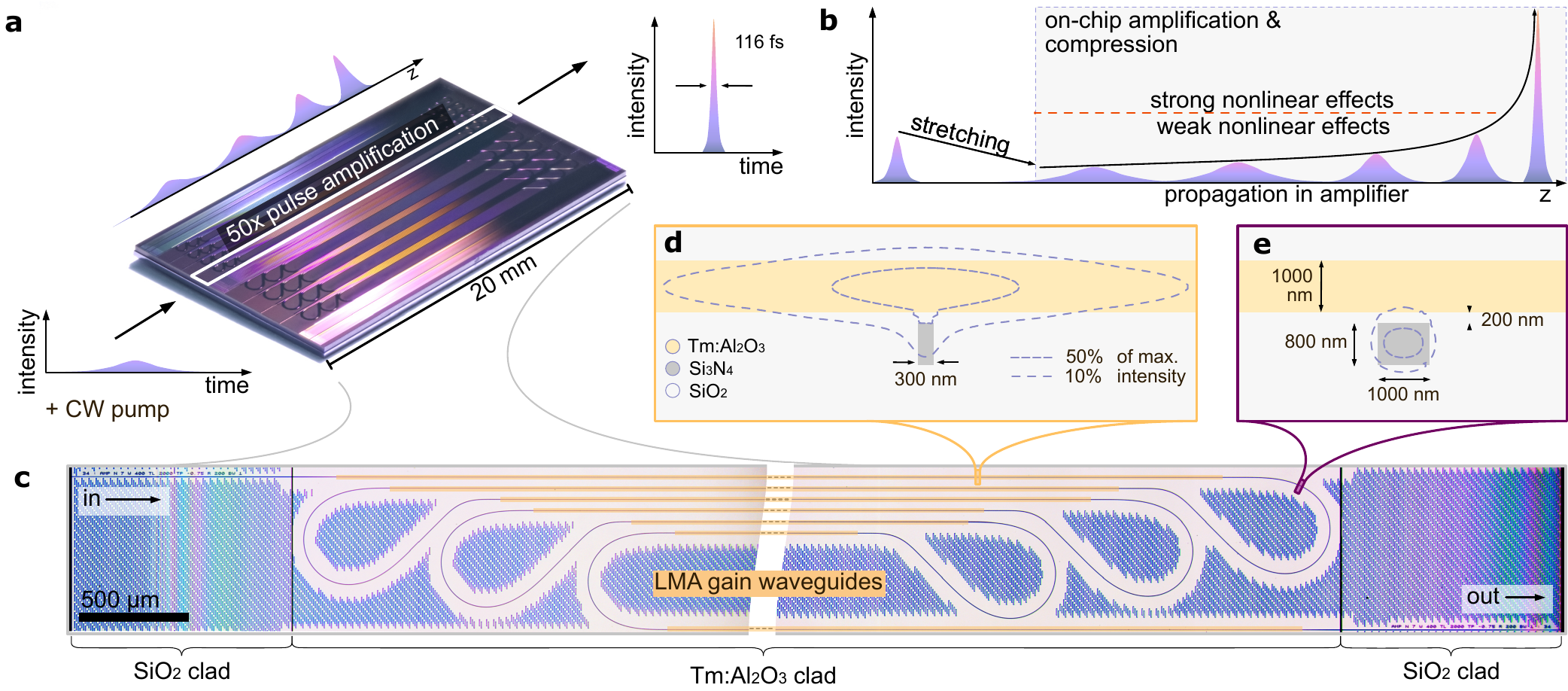} 
  \caption{
      \textbf{Ultrafast pulse amplification on a chip}.
        \textbf{a}, Photograph of the chip, hosting several amplifiers. Low energy chirped pulses are amplified 50-fold. The output pulse is a high-peak power, nearly time-bandwidth-limited femtosecond pulse.
         \textbf{b}, After temporally stretching (chirping) the input pulse, it is coupled to the amplifier chip, where it is gradually amplified and compressed, avoiding strong nonlinearities  until passing the output section.
        \textbf{c}, Composite photograph of an amplifier structure, consisting of large mode-area (LMA) gain waveguides, folded into a small 15~mm$^2$-scale footprint. 
        The waveguides are defined in a silicon nitride (Si$_3$N$_4$) layer, where within the area of the gain waveguides, an active thulium-doped alumina layer (Tm$^{3+}$:Al$_2$O$_3$) is used as a top cladding; in the input and output sections a silica (SiO$_2$) cladding is used. \textbf{d}, Layer structure in the LMA gain waveguide section. Intensity contours of the optical mode illustrate that most of the power is propagating in the gain layer for optimal amplification. \textbf{e}, Same as (d) but for the bent waveguide sections. The mode is strongly confined to the Si$_3$N$_4$-core, enabling low loss propagation through the bends.
    }
  \label{fig:intro}
\end{figure*}

Previous work has already demonstrated chip-integrated (quasi-) continuous-wave amplifiers based on rare-earth-doped waveguides \cite{agazzi_monolithic_2010, vazquez-cordova_erbium-doped_2014, sun_giant_2017, ronn_ultra-high_2019, kiani_thulium-doped_2019, frankis_erbium-doped_2020, mu_high-gain_2020, ronn_erbium-doped_2020, liu_photonic_2022, jia_integrated_2022} sustaining power levels up to 1~W \cite{singh_watt-class_2023}, heterogeneous semiconductor integration \cite{davenport_heterogeneous_2016, davenport_integrated_2018, van_gasse_27_2019, op_de_beeck_heterogeneous_2020} or nonlinear parametric gain \cite{choi_optical_2021, ledezma_intense_2022, riemensberger_photonic_2022}. However, the amplification of femtosecond pulses in an integrated photonics setting has remained challenging, with nonlinear pulse distortion already at low peak power levels of 20~W \cite{liu_photonic_2022}. Indeed, the strong light confinement in integrated waveguides results in strong nonlinearities, typically 1000-times higher than those in optical fibers. While often advantageous \cite{leuthold_nonlinear_2010}, e.g. for broadband supercontinua \cite{ dudley_supercontinuum_2006, lamont_supercontinuum_2008, kuyken_mid-infrared_2011, halir_ultrabroadband_2012}, nonlinear optical effects during the amplification process can lead to the irreversible degradation of the pulse within sub-mm propagation distances. 

Tabletop laser systems have historically faced a similar challenge. Nonlinear optical effects would prevent the amplification of pulses 
%with preserved shape \cite{dudley_self-similarity_2007-1}, 
or even result in damage of the amplifier itself \cite{wood_laser_1975}. A major breakthrough came through the demonstration of \textit{chirped pulse amplification} (CPA) \cite{strickland_compression_1985}. In CPA the pulse is temporally stretched (\textit{chirped}) to lower its peak power, amplified, and then temporally re-compressed only after amplification. While stretching can be readily accomplished at low power levels through a dispersive optical element, amplification and compression require advanced optical setups to manage nonlinearity, group velocity dispersion (GVD), and for ultra-short pulses, also higher-order dispersion.

Here, we demonstrate femtosecond pulse amplification on an integrated photonic chip to hundreds of Watts of peak power (Fig.~\ref{fig:intro}a), a previously lacking cornerstone for chip-scale femtosecond technology.
Our approach pursues a concept that resembles CPA to mitigate the strong nonlinear optical effects that are usually associated with integrated photonics. Pre-chirped input pulses propagating through the amplifier are simultaneously amplified and temporally compressed by managing nonlinearity and dispersion so that ultrashort pulse duration and high-pulse peak power are reached at the amplifier's output facet (Fig.~\ref{fig:intro}b). In this way, we achieve $>$50-fold amplification of ultrashort pulses from a 1~GHz femtosecond source (center wavelength 1815~nm) to 800~W peak power and a final pulse duration of 116~fs. Our experimental results are in excellent agreement with a numerical pulse propagation model that includes the dynamics of the Tm-doped gain medium.
%These results demonstrate experimentally and numerically that femtosecond pulse duration and high peak power on chip are compatible, opening new opportunities for ultrafast integrated photonics. 

\begin{figure*}[t!]
  \centering
  \includegraphics[width=2\columnwidth]{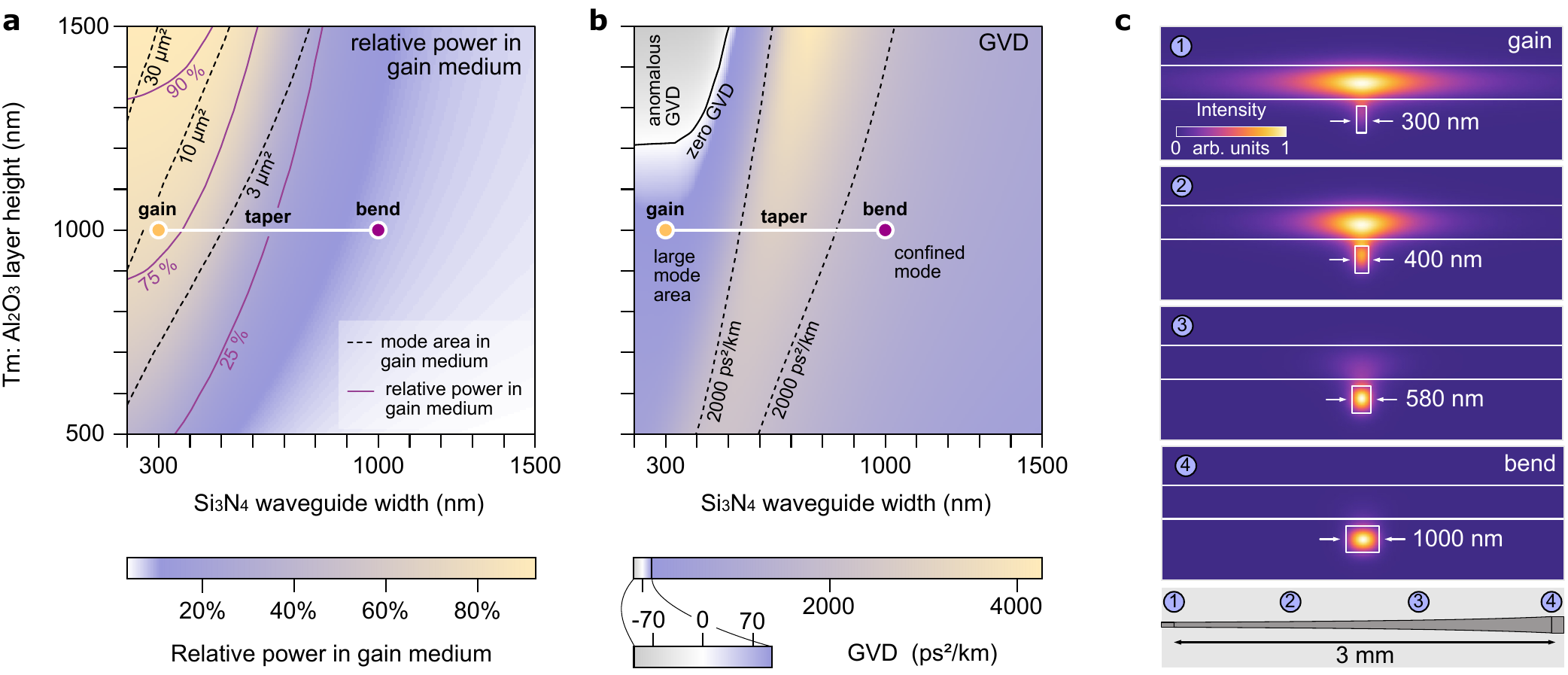} 
  \caption{
      \textbf{Amplifier design}.
        \textbf{a}, Relative power fraction in the transverse magnetic (TM) mode that is contained in the Tm$^{3+}$:Al$_2$O$_3$ gain layer (colorscale, purple contours) and mode-area in the in the gain layer (dashed black contours) in dependence of the Tm$^{3+}$:Al$_2$O$_3$ layer height and the Si$_3$N$_4$ waveguide width; cf. Methods for definition of the mode-area. The locations of the gain waveguides (``gain"), bends (``bend") and tapers (``taper") in the parameter space are indicated. 
         \textbf{b}, Group velocity dispersion (GVD) (colorscale, dashed black contours); axes and labels as in (a). 
        \textbf{c}, Simulated intensity profiles for different Si$_3$N$_4$ waveguide widths as they occur in gain, bent and tapered waveguide sections. Their positions along the adiabatic taper-profile are indicated. The simulations are based on a finite-element method model.
    }
  \label{fig:FEM_sim}
\end{figure*}

\begin{figure*}[t!]
  \centering
  \includegraphics[width=2\columnwidth]{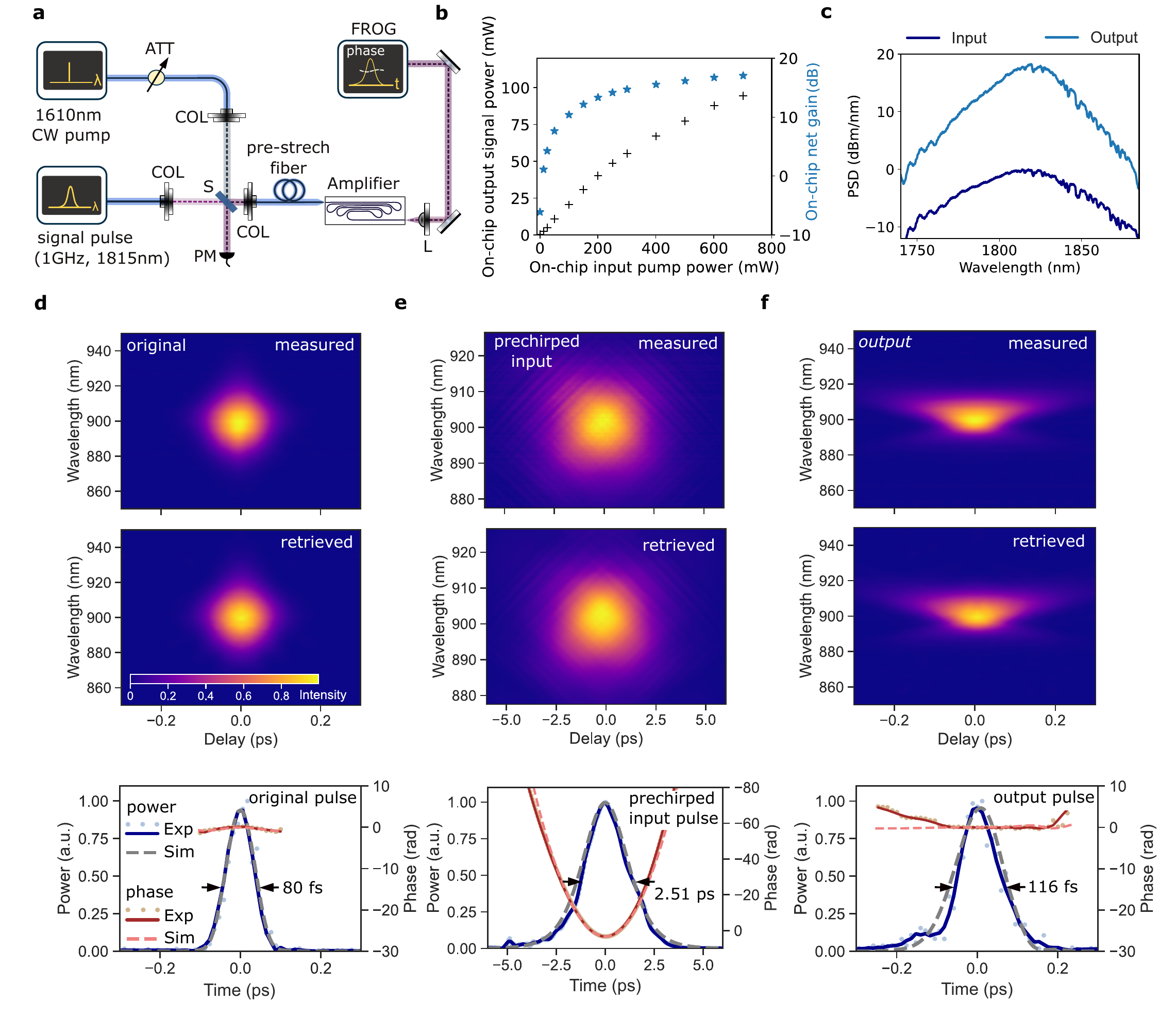}
  \caption{
      \textbf{Experiments}. \textbf{a}, Experimental setup. CW: continuous-wave; ATT: attenuator; COL: collimator; S: beam-splitter; L: lens; PM: power-meter
      \textbf{b}, Amplifier output power and on-chip net gain as function of the on-chip pump power.
      \textbf{c}, Optical spectra before and after amplification; PSD: power spectral density.
      \textbf{d,e,f}, Frequency resolved optical gating (FROG) traces showing second harmonic intensity as function of FROG delay (256x256 data points) and reconstructed temporal pulse (power and phase) of the original, the pre-chirped input and the output pulses. Dots and solid lines represent measured and smoothed measured data, dashed lines represent simulated data. The respective FROG errors are 1.282$\cdot$10$^{-5}$, 2.75$\cdot$10$^{-6}$, and 1.19$\cdot$10$^{-5}$. 
    }
  \label{fig:experiments}
\end{figure*}

\begin{figure}[t!]
  \centering
  \includegraphics[width=1\columnwidth]{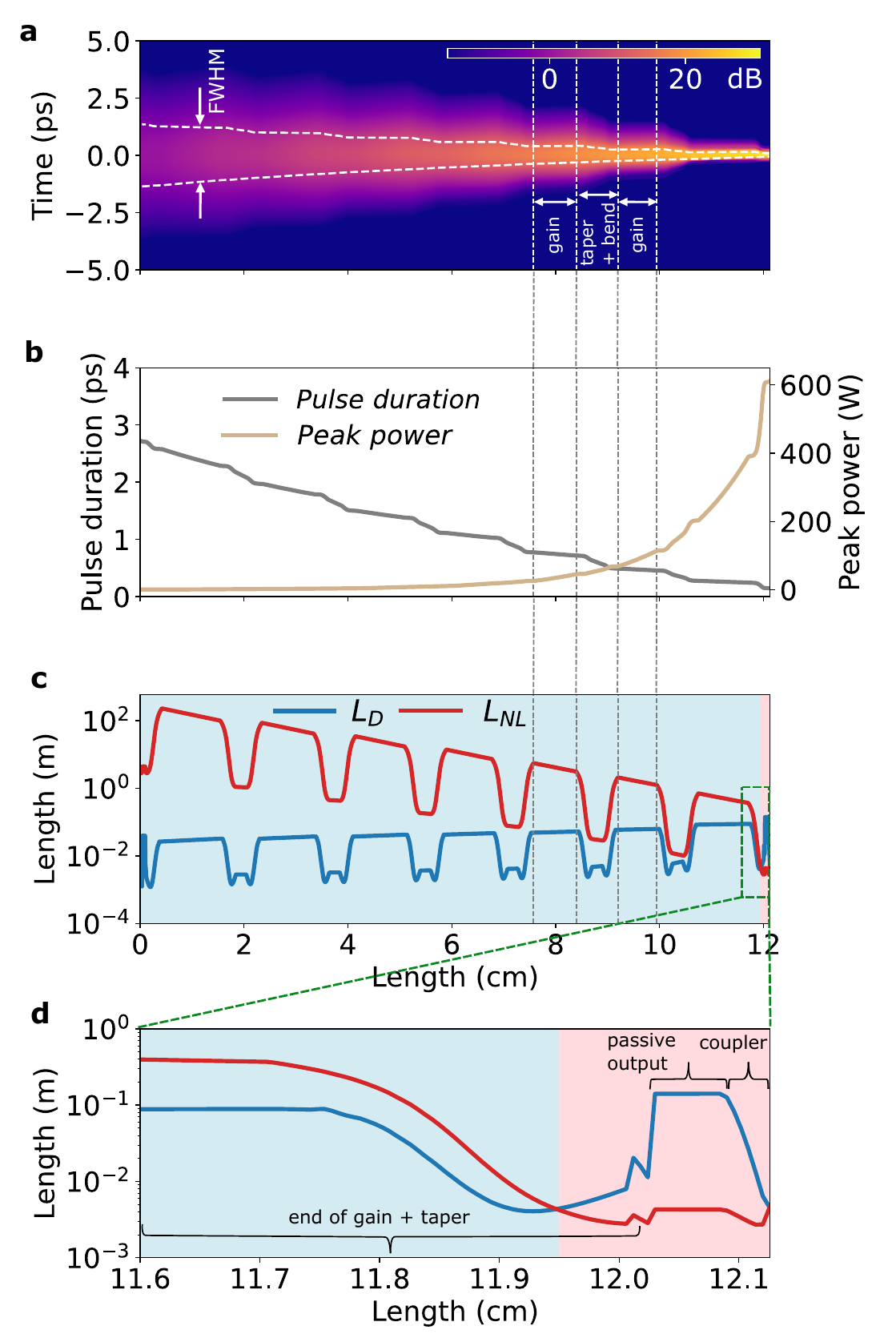}
  \caption{
      \textbf{Numerical simulation}. \textbf{a}, Evolution of the temporal pulse power while propagating through the amplifier chip in a co-moving reference frame. One sequence of the alternating gain, bend and taper sections is indicated. The contours indicate the full-width-half-maximum (FWHM) pulse duration.
      \textbf{b}, Pulse duration and pulse peak power while propagating through the amplifier chip. \textbf{c}, Evolution of the pulse's dispersion length $L_\mathrm{D}$ and nonlinear length $L_\mathrm{NL}$ while propagating through the amplifier chip. The blue background color highlights where the propagation is dominated by linear optical effects ($L_\mathrm{D}<L_\mathrm{NL}$); nonlinear optical effects dominate only in the last $1.5$~mm of the entire $>$12~cm long propagation distance (red background color). \textbf{d}, Magnified portion of panel~c, where the gain section and taper to the output waveguide, the passive output waveguide, and the inverse taper for chip-output coupling are indicated. See Methods for details on the simulation.
    }
  \label{fig:simulation}
\end{figure}

\subsection*{Results}
The amplifier (Fig.~\ref{fig:intro}c) is fabricated in a CMOS-compatible, scalable silicon nitride (Si$_3$N$_4$) photonic-chip platform (here with a 800~nm thick waveguide layer).
The total length of the amplifier waveguide structure is 12~cm, which is integrated in a 15~mm$^2$-scale footprint. It comprises a total of $\sim$10~cm of straight waveguide sections that provide gain and curved waveguide sections for compact integration. To achieve on-chip gain, the silica (SiO$_2$) top cladding is selectively removed in the amplifier section down to 200~nm above the Si$_3$N$_4$ layer. In a next step, a 1000~nm thick thulium-doped alumina (Tm$^{3+}$:Al$_2$O$_3$) gain layer is deposited. The refractive index of the gain layer $n\approx1.72$ is between that of Si$_3$N$_4$ and  SiO$_2$. The estimated Tm concentration is 3.5$\cdot$10$^{20}$cm$^{-3}$.
A 1000~nm-thick silica layer is deposited on top for protection (see Methods for more details on fabrication). 
In the straight gain sections the width of the Si$_3$N$_4$ waveguide is reduced to 300~nm, resulting in a single-mode waveguide with large mode-area that is mostly contained in the active alumina gain layer (Fig.~\ref{fig:intro}d). Such large mode area (LMA) gain waveguides \cite{singh_towards_2020} can provide high-gain per unit length, low-nonlinearity, and high saturation power \cite{singh_watt-class_2023}, and have already enabled Q-switched lasers with nanosecond pulse duration \cite{shtyrkova_integrated_2019, singh_silicon_2024}.
In the curved sections the waveguide width is 1000~nm for strong mode confinement (Fig.~\ref{fig:intro}e) and low-loss propagation; Euler-bends with a minimal radius of 200~µm guarantee an adiabatic low-loss transition to the bends \cite{vogelbacher_analysis_2019}.
  
The amplifying and bending waveguide sections are connected by straight adiabatically tapered waveguides. To increase mode confinement for low reflections and scattering at the cladding material transition, input and output waveguides are 2000~nm wide. Inverse tapers at the facets of the chip enable low-loss coupling from and to optical fibers or free-space.
The footprint of the entire amplifier structure is below 15~mm$^2$. 
The choice of Tm-doping is motivated by its large gain bandwidth \cite{su_ultra-compact_2016, li_high-power_2017, shtyrkova_integrated_2019, kiani_thulium-doped_2019, singh_silicon_2024} and its importance for emerging applications such as laser ranging, free-space communication \cite{li_thulium-doped_2013} and mid-infrared supercontinua \cite{luo_mid-ir_2016}; however, the amplifier platform can also support erbium-doping for operation in the telecommunication wavelength window \cite{Dawson_2022}, and likely 
other rare-earth dopants such as neodymium and ytterbium for operation around 1~$\mu$m wavelength.

The design of the geometry of the amplifier is based on numerical simulation of key waveguide parameters obtained through a finite element electro-magnetic mode solver. For these simulations to be accurate, we utilize wavelength dependent refractive index data that was obtained for all involved materials through broadband ellipsometry. Fig.~\ref{fig:FEM_sim}a,b show, for the fundamental transverse magnetic mode (TM), the relative optical power in the gain layer and the GVD as a function of the Si$_3$N$_4$ waveguide width and the Tm$^{3+}$:Al$_2$O$_3$ layer height. To achieve high gain and high saturation power, a large power fraction of the optical mode in the Tm$^{3+}$:Al$_2$O$_3$ layer is generally desirable. This also reduces the impact of nonlinear optical effects by (i) lowering the intensity through a large mode-area, (ii) a 10-fold reduced nonlinear material index of the gain layer compared to the silicon nitride waveguide, and (iii) by achieving high-gain amplification over short distances. However, a too large mode-area (and the correspondingly high number of excited ions) would increase the minimum signal input power required to avoid excessive spontaneous emission or parasitic lasing, particularly in a co-propagating (forward) pumping scheme.
In addition to a high power fraction in the gain medium, we aim at an all-normal GVD along the entire amplifier structure to achieve monotonous temporal pulse compression and robustness against pulse breakup by nonlinear modulation instability (MI) \cite{dudley_instabilities_2014}.
As Fig.~\ref{fig:FEM_sim}a and b show, it is indeed possible to simultaneously fulfill these criteria for the gain-waveguide, taper and bend for a Tm$^{3+}$:Al$_2$O$_3$ layer thickness below 1200~nm. Here, we use a chip with a layer thickness of 1000~nm in the normal GVD regime.

The location of the gain waveguide, bends and connecting tapers in the design parameter space are indicated in Fig.~\ref{fig:FEM_sim}a,b and their mode profiles are shown in Fig.~\ref{fig:FEM_sim}c. In this design the effective mode-area in the gain layer is $A^\mathrm{gain}_\mathrm{eff}\approx7$~µm$^2$ with an effective nonlinear parameter of $\gamma=0.007$~W$^{-1}$m$^{-1}$ (in the 1000~nm wide waveguide $\gamma=0.7$~W$^{-1}$m$^{-1}$) (see Methods).
The group-delay dispersion (GDD, i.e. the length-integrated GVD) of the amplifier chip is 7.79$\cdot$10$^{-26}$~s$^2$, which is large enough to allow pre-chirping to pico-second duration pulses. For ultrashort pulses, besides second order dispersion (GVD), third order disperison can be important. As we show in the Supplementary Information (SI, Section~1, Figure~S1), the distinct third order dispersion characteristics of bends and tapers can be leveraged to minimize detrimental impacts of third order dispersion.

The experimental setup for testing the amplifier is shown in Fig.~\ref{fig:experiments}a, including the input signal pulse and the 1610~nm continuous-wave (CW) pump sources that are externally combined and coupled to the chip. 
For convenience, the pump source is here implemented via a diode laser that is amplified in an erbium-doped fiber amplifier. Future integration may utilize high power single-mode laser diodes, which are commercially available with power levels of larger than 400~mW, and which could be combined in a polarization multiplexed and/or bi-directional pumping configuration. Alternatively, 790~nm pump diodes may be used to excite the gain medium via power-efficient cross-relaxation processes; given the significance of Tm-doped gain materials, we anticipate high-power pump diodes to become readily available.
The output of the amplifier is collected via free-space optics to avoid additional dispersive or nonlinear optical effects and characterized via frequency resolved optical gating (FROG) \cite{trebino_measuring_1997}. 
As input pulse source we use a femtosecond laser source (cf. Methods) providing close to bandwidth-limited pulses of 80~fs duration at a central wavelength of 1815~nm and a pulse repetition rate of 1~GHz of which we couple 1.81~mW of average power ($1.81$~pJ pulse energy) to the chip.
Prior to coupling the pulses to the chip, they are pre-chirped in a short stretch of standard anomalous dispersion optical fiber. The fiber can easily be adjusted in its length to explore different levels of pre-chirping. However, we emphasize that such pre-chirping can also be implemented on-chip without suffering from nonlinear effects owing to the low pulse energy prior to amplification. Relevant approaches include strongly anomalous dispersion waveguides \cite{yu_integrated_2022}, 
Bragg-gratings \cite{du_silicon_2020, sahin_large_2017, li_large_2023}, or coupled gratings \cite{choi_high_2021, tan_monolithic_2010}.

As an aside, we note that it is in principle also possible to send the pulses directly, without pre-chirping, into the amplifier and to perform re-compression after the amplification (SI, Section~2, Fig.~S2). 
However, this would require dispersion compensation at higher power levels after amplification, which, in contrast to pre-chirping at lower power, is challenging due to non-trivial nonlinear compression dynamics. If carefully optimized, self-similar nonlinear pulse amplification in the normal dispersion regime \cite{fermann_self-similar_2000, dudley_self-similarity_2007-1} , followed by anomalous dispersion and/or nonlinear pulse compression may results in pulse shortening; however, this is not pursued here and only works for very specific input pulse parameters.

Fig.~\ref{fig:experiments}b shows the average power of the amplified pulses as well as the on-chip net gain for different levels of pump power. For the highest on-chip pump power of 700~mW, we observe an on-chip output signal power of 95~mW and a maximal gain of 17~dB, i.e. a factor of 52 (see Methods for details on calibration); the optical spectrum before and after amplification is shown in Fig.~\ref{fig:experiments}c.
Next, we vary the length of the pre-chirping fiber and measure the temporal characteristics of the output pulses directly after the chip via FROG. For an optimal pre-chirping fiber length of 146.5~cm, commensurate with the opposite amount of total dispersion along the amplifier chip, we obtain an amplified and compressed pulse directly at the output facet of the chip of 116~fs duration. The temporal characterization of the original, pre-chirped and amplified pulses are shown in Fig.~\ref{fig:experiments}d,e and f, respectively. 
The pre-chirped pulses have a duration of $2.5$~ps and a markedly quadratic temporal phase. In contrast, the output pulses of the amplifier are 116~fs in duration and exhibit a largely flat temporal phase indicating that they are close to time-bandwidth-limited. The asymmetric low intensity tail of the reconstructed pulse (and a `butterfly' shape in the FROG trace) hints at a weak residual third order dispersion.
In total, this experiment signifies the successful more than 50-fold amplification of a femtosecond pulse in an integrated photonic chip.  
The obtained on-chip pulse energy of $95$~pJ and peak power of $>800$~W are well-suited for photonic chip-based spectral generation from infrared to ultraviolet wavelength \cite{wu_visible_2023, ludwig_ultraviolet_2023} or self-referencing \cite{johnson2015} for absolute optical metrology. 
In integrated microwave photonics \cite{marpaung_integrated_2019}, ultrashort pulses could improve shot-noise limited performance by several orders of magnitude \cite{quinlan_exploiting_2013} with impact on navigation and communication applications.
If higher average output power is needed, a longer amplifier structure and larger mode-area gain waveguides can be used.
We note that the final output pulses of 116~fs duration are longer than the 80~fs input pulse. This correlates well with their respective spectral bandwidth of 5.1~THz and 3.5~THz, respectively. We therefore attribute the increase in pulse duration to a finite gain bandwidth, but also to residual third order dispersion. Potentially, although less likely, nonlinear spectral compression could play a role \cite{boscolo_impact_2018}.

Complementing the experiment, we perform numerical simulations of on-chip pulse amplification based on the nonlinear Schrödinger equation (NLSE) combined with a thulium gain model (see Methods). Using the experimental parameters of the input pulse, pump power and waveguide parameters, we simulate the pulse amplification and compare it with the experimental data in Fig.~\ref{fig:experiments}d,e,f (bottom). The simulation is in excellent agreement with the experimental observations and we can use it to gain further insights into the pulse propagation during the amplification process. Fig.~\ref{fig:simulation}a shows the simulated temporal dynamics of amplification and pulse shortening. We extract the pulse duration as a function of propagation distance and confirm the intended monotonous pulse compression that permits to keep the pulse peak power $P_\mathrm{p}$ and the impact of nonlinear effects low throughout most of the amplifier length (Fig.~\ref{fig:simulation}b). To illustrate this point, we plot in Fig.~\ref{fig:simulation}c the nonlinear length $L_\mathrm{NL}=(\gamma P_\mathrm{p})^{-1}$, 
the characteristic propagation length scale on which nonlinear effects become relevant. It can be seen that $L_\mathrm{NL}$ is consistently longer than the stretches of propagation to which it applies, indicating the limited impact of nonlinear effects. We also indicate the characteristic length scale of dispersive effects 
$L_\mathrm{D}=\mathrm{TBP}^2/(\Delta f^2|\mathrm{GVD}|)$, which is significantly shorter than $L_\mathrm{NL}$ for most of the propagation, confirming that pulse propagation is dominated by linear optical effects ($\Delta f$ is the spectral bandwidth of the pulse as obtained from the simulation data at each point of the propagation and $\mathrm{TBP}$ is the time-bandwidth product, here set to 0.315).
\\

\subsection*{Discussion}
In summary, we have demonstrated ultrashort femtosecond pulse amplification in an integrated photonic chip.  
The central challenge of strong optical nonlinearity inherent to integrated photonics is effectively addressed through tailored large mode-area gain waveguides and the design of the waveguide's dispersion, permitting stable pulse propagation and compression, including in the tightly confining bends. The achieved 50-fold on-chip amplification leads to a peak power of 800~W. Longer amplifying waveguides and larger mode-area can support even higher average output power, accommodating higher pulse repetition rates. As we discuss in the SI (Section~3, Fig.~S3), substantially increasing the peak power while avoiding nonlinear effects would require a design with increased waveguide dispersion, or shorter pulses which are, in principle, supported by the Tm-doped gain medium. Future work, may also  leverage deliberate nonlinear effects such as self-similar pulse amplification in the normal dispersion regime \cite{fermann_self-similar_2000, dudley_self-similarity_2007-1} followed by anomalous dispersion and potentially nonlinear compression to provide high-power few-cycle pulses \cite{carlson_generating_2019}. As such, our results open new opportunities for scalable high-power fully-chip integrated ultrafast sources. More broadly, they bring femtosecond technology with pulse peak-power otherwise only attainable in table-top systems to the chip-level.

\hspace{3cm}
%================================
% Methods
%================================
\subsection*{Methods}
\small
% \footnotesize
\paragraph{Sample fabrication.} 
The passive Si$_3$N$_4$ chips were fabricated by Ligentec SA using UV optical lithography, based on a 800~nm-thick Si$_3$N$_4$ waveguide platform embedded in a silica cladding. In the area of the gain waveguides, the top silica cladding is removed (a 200~nm silica spacer layer remains). A Tm$^{3+}$:~Al$_2$O$_3$ gain layer with a thickness of 1000~nm is then deposited in the local opening using radio frequency sputtering technique \cite{van_emmerik_relative_2020} with an estimated Tm$^{3+}$ ion concentration of 3.5$\cdot$10$^{20}$cm$^{-3}$. The gain layer is cladded with a 1000~nm-thick silica layer on top for protection. 
Finally, the chip is annealed at 500~\textdegree C for 6 hours to reduce the OH and water absorption in the cladding layer.

\paragraph{Nonlinear coefficient and effective mode-area.}
The nonlinear coefficient $\gamma$ of the waveguides is \cite{koos_nonlinear_2007}  
\begin{align*}
    \gamma = \frac{\epsilon_0 \omega_s}{\mu_0 c}\frac{\iint n^2(x,y) n_2(x,y) |\mathbf{E(x,y)}|^4 \mathrm{d}x\mathrm{d}y}{\left|\iint \mathrm{Re}\left[\mathbf{E}(x,y) \times \mathbf{H}^*(x,y)\right]_z \mathrm{d}x\mathrm{d}y\right|^2}
\end{align*}
where the integral extends over the entire heterogeneous material geometry. $\mathbf{E}$ and $\mathbf{H}$ are the complex electric and magnetic field vectors, $x$ and $y$ the transverse spatial coordinates.  
The effective mode-area $A_\mathrm{eff}^\mathrm{gain}$ in the homogeneous gain medium is 
\begin{align*}
    A^\mathrm{gain}_\mathrm{eff} = \frac{\left(\iint_\mathrm{gain}|\mathbf{E}(x,y)|^2\mathrm{d}x\mathrm{d}y\right)^2}{\iint_\mathrm{gain}|\mathbf{E}(x,y)|^4\mathrm{d}x\mathrm{d}y}
\end{align*}
where the integration is restricted to the Tm$^{3+}$:Al$_2$O$_3$ gain layer.

\paragraph{Ultrafast signal source.}
150~fs, 1~GHz pulses at 1560~nm are launched into a highly nonlinear fiber (HNLF) to generate a Raman-shifted soliton pulse centered at a wavelength of 1815~nm. A free-space 1700~nm long pass filter (LPF) is used to remove residual short wavelengths contributions.

\paragraph{Power calibration and gain measurements.}
Pump and signal fiber-to-chip coupling losses are measured using passive (no gain layer) waveguides and a symmetric lensed fibers configuration for input and output coupling. Measured pump and signal coupling losses are 2.9 and 4.2~dB per facet, respectively. We extract the net gain by directly comparing the output and input signal powers after calibrations and after filtering out the transmitted pump using a free-space long-pass filter.

%\paragraph{Numeric simulation}
\paragraph{Model for the amplification dynamics.}
Based on the NLSE, our model for propagation and amplification of the signal pulses combines effects of nonlinearity, gain and dispersion. The optical nonlinearity is treated in the time domain
\begin{equation*}
\begin{aligned}\label{NLSE1}
\frac{\partial A_k(z)}{\partial z}=i \gamma \lvert A_k(z) \rvert^2 A_k(z),
\end{aligned}
\end{equation*}
where $A_k(z)$ is the complex temporal field amplitude of the signal, defined in the co-moving time frame on the discrete temporal points $T_k$ ($k=0, 1, 2, .., N-1$) and normalized such that $\sum_k \lvert A_k(z) \rvert^2\, \Delta T$ is the pulse energy at position $z$ with $\Delta T = T_{k+1}-T_k$. Dispersion and gain are treated in the frequency domain
\begin{equation*}\label{NLSE2}
\begin{aligned}
\frac{\partial \tilde{A}_j(z)}{\partial z}=\frac{g_{S,j}(z)}{2}\tilde{A}_j(z)+i\sum_{n\geq2} \frac{\beta_n}{n!}(\omega_j-\omega_S)^n\tilde{A}_j(z)
\end{aligned}
\end{equation*}
where $\tilde{A}_j(z)$ is the discrete Fourier-transform of $A_k(z)$, with frequencies $\omega_j$ ($j=0, 1, 2, .., N-1$), and $\omega_S$ is the signal's center frequency. The frequency resolution is $\Delta\omega = \frac{2\pi}{N\Delta T}$, the signal gain is $g_{S,j}(z)$, and the $\beta_n$ describe the dispersion.

We model the optical gain by accounting for the populations in the first three lower-lying levels ($^3H_6$, $^3F_4$ and $^3H_5$). The gain model also accounts for the fractions $f_{a}$ and $f_{q}$ of active and quenched ions ($f_a + f_q = 1$), respectively. Assuming a steady state for the gain medium, the rate equations including energy transfer upconversion (ETU) process for 1610~nm pumping are given by \cite{agazzi2013EnergyTransferUpconversionModelsTheir, vazquez-cordova_erbium-doped_2014}: 

\begin{align*}%\label{rate}
\begin{split}
\frac{dN_{2,a/q}(z)}{dt} = & W_{ETU} N^2_{1,a/q}(z)-\frac{1}{\tau_2}N_{2,a/q}(z)=0\\
\frac{dN_{1,a/q}(z)}{dt}= & R_{P,a/q}(z)-R_{S,a/q}(z)-W_{ETU} N^2_{1,a/q}(z)\\
& +\frac{1}{\tau_2}N_{2,a/q}(z)-\frac{1}{\tau_{1,a/q}}N_{1,a/q}(z)=0\\
f_{a/q}(z)N_d = & N_{0,a/q}(z)+N_{1,a/q}(z)+N_{2,a/q}(z)\\
R_{P,a/q}(z) = & \frac{1}{\hbar\omega_P}\frac{P_P(z)}{A_\mathrm{eff}(\omega_P)}\\ & \times \Bigr[\sigma_{a}(\omega_P) N_{0,a/q}(z)-\sigma_{e}(\omega_P) N_{1,a/q}(z)\Bigr]\\
R_{S,a/q}(z) = & \sum_j\frac{1}{\hbar\omega_S}\frac{P_{S,j}(z)}{A_\mathrm{eff}(\omega_S)} \\ & \times\Bigr[\sigma_{e}(\omega_j)N_{1,a/q}(z)-\sigma_{a}(\omega_j) N_{0,a/q}(z)\Bigr] 
 \makeatletter 
\renewcommand{\theequation}{S\@arabic\c@equation}
\makeatother
\end{split}
\end{align*}\\ where the concentrations $N_{0,1,2}$ describes the populations in the levels $^3H_6$, $^3F_4$ and $^3H_5$, respectively, the subscripts $a$/$q$ denote the active or quenched ions and $N_d$ is the total ion concentration; $\tau_i$ is the luminescence lifetime of level $i$ and $W_{ETU}$ is a parameter accounting for the ETU processes $^3F_4$, $^3F_4 \rightarrow$ $^3H_5$, $^3H_6$ \cite{vazquez-cordova_erbium-doped_2014, loiko_stochastic_2016}. $R_{p,a/q}$ and $R_{s,a/q}$ are the pump and signal rates. The pump power $P_P$ is monochromatic with frequency $\omega_P$ and the signal power contained in the spectral discretization intervals is 
$P_{S,j}(z)=\frac{\Delta T^2}{T}|\tilde{A}_j(z)|^2 \frac{\Delta \omega}{2\pi}= N^{-2}|\tilde{A}_j(z)|^2$.
For the pump and signal intensities, we neglect the transverse spatial intensity profiles (orthogonal to the propagation direction) and instead approximate them via effective mode areas
$A_\mathrm{eff}(\omega_P)\approx A_\mathrm{eff}(\omega_S)\approx$ \SI{7.5}{\micro m^2}. 
Further, $\sigma_{e}$ and $\sigma_{a}$ are the frequency dependent effective emission and absorption cross-sections and $\hbar$ is the reduced Planck constant. 
The propagation equations of the pump and signal power, which describe the gain along the propagation direction $z$ are
\begin{align*}
\frac{1}{P_P (z)}\frac{dP_P (z)}{dz}= & \Gamma_P \Bigr[\sigma_{e}(\omega_P) \Bigl(N_{2,a}(z)+N_{2,q}(z)\Bigl) \\
& -\sigma_{a}(\omega_P)\Bigl(N_{0,a}(z)+N_{0,q}(z)\Bigl)\Bigr] -\alpha
\end{align*}
and
\begin{align*}\label{Ps}
\frac{1}{P_{S,j}(z)}\frac{dP_{S,j} (z)}{dz}= & \Gamma_S \Bigr[\sigma_{e}(\omega_j)\Bigl(N_{1,a}(z)+N_{1,q}(z)\Bigl) \\
& -\sigma_{s,a}(\omega_j)\Bigl(N_{0,a}(z)+N_{0,q}(z)\Bigl)\Bigr] -\alpha .
\end{align*}
Here, $\Gamma_{S}$ and $\Gamma_{P}$ are the power fractions of signal and pump in the gain medium, respectively, and $\alpha$ describes the propagation loss which we assume to be the same for both pump and signal. While the pump propagation equation is used directly to model the evolution of $P_P(z)$, the evolution of the signal is modeled via its field amplitude as described above using $g_{S,j}(z)=\frac{1}{P_{S,j}(z)}\frac{dP_{S,j} (z)}{dz}$.

% Finally we obtain
% \begin{align*}
%     g_{s}(\omega_k, z) = \frac{1}{P_{s} (\omega_k,z)}\frac{dP_{s} (\omega_k, z)}{dz}\, .
% \end{align*}
%, while $N_{i}(z)$ is the population density of level $i$= 0, 1 and 2.

% The gain $g_s(z,\omega)=1/P_s(z,\omega)\cdot dP_s(z,\omega)/dz$ in every propagation step $dz$ is then updated in the NLSE. Here, we considered only the signal propagation in the NLSE, as the pump is a continuous wave, therefore neither dispersion nor nonlinearities are expected from the pump. The NLSE, including the signal gain $g_{s}$, is written as:

% \begin{equation*}\label{NLSE}
% \begin{aligned}
% \frac{\partial A_{s}(T,z)}{\partial z}=\frac{g_{s}(z)}{2}A_{s}+\sum_{n=2}^N (i)^{n+1}\frac{\beta_n}{n!}\frac{\partial^n A_{s}}{\partial T^{ n}}+i \gamma \lvert A \rvert^2 A_{s},
% \end{aligned}
% \end{equation*}Where  $\beta_n$=$\partial^n \beta/\partial \omega^n$ are the dispersion coefficients associated with the Taylor series expansion of the dispersion function $\beta(\omega)$ at $\omega=0$ and $T=t-z/v_{s}$ is the retarded time.

% \subsection*{Data availability}
% \small
% The data shown in the plots have been deposited in the Zenodo database https://zenodo.org/records/12968259; DOI: 10.5281/zenodo.12968259.

%================================
% Acknowledgments
%================================
\subsection*{Acknowledgments}
\small
This project has received funding from Horizon 2020 research and innovation program (FEMTOCHIP, grant agreement No 965124 (M.G., F.X.K., S.M.G.B., T.H.)), from the European Research Council (ERC) under the EU’s Horizon 2020 research and innovation program (STARCHIP, grant agreement No 853564 (T.H.); CounterLight, grant agreement No 756966(P.D.H.)), from the Deutsche Forschungsgemeinschaft (SP2111, contract number 403188360 (F.X.K.)), and through the Helmholtz Young Investigators Group VH-NG-1404 (T.H.); the work was supported through the Maxwell computational resources operated at DESY.

%================================
% Author contribution
%================================
\subsection*{Author Contributions}
M.A.G., T.W., T.H. designed the amplifier chip. M.A.G. developed and performed tests for gain waveguide development. M.A.G., M.L. designed and built the setup for pulse characterization, performed pulse amplification experiments and analyzed the data. T.V. contributed to the development of pulse input source. M.A.G. developed and performed the numerical simulations.  H.F., M.G. fabricated the silicon nitride chip. K.W., S.M.G.B. developed and deposited the gain medium. S.Z., T.B., P.D.H. supported cladding development. M.S., J.L., N.S., F.X.K. contributed to the development of gain medium and amplifier. T.H. supervised the work. M.A.G., T.H. prepared the manuscript with input from all authors.

%================================
% Disclosures
%================================

\subsection*{Competing interests}
\small
We declare that none of the authors have competing interests.

%================================
% References
%================================

% \printbibliography
\bibliography{Amplifiers_v3}

%================================
% SI
%================================

\newpage
\onecolumn

\setcounter{page}{1}
\setcounter{figure}{0}

\title{Supplementary Information - \\ Femtosecond pulse amplification on a chip}

\maketitle

\section{Third order dispersion}
\begin{figure}[h!]
  \centering
\includegraphics[width=0.5\columnwidth]{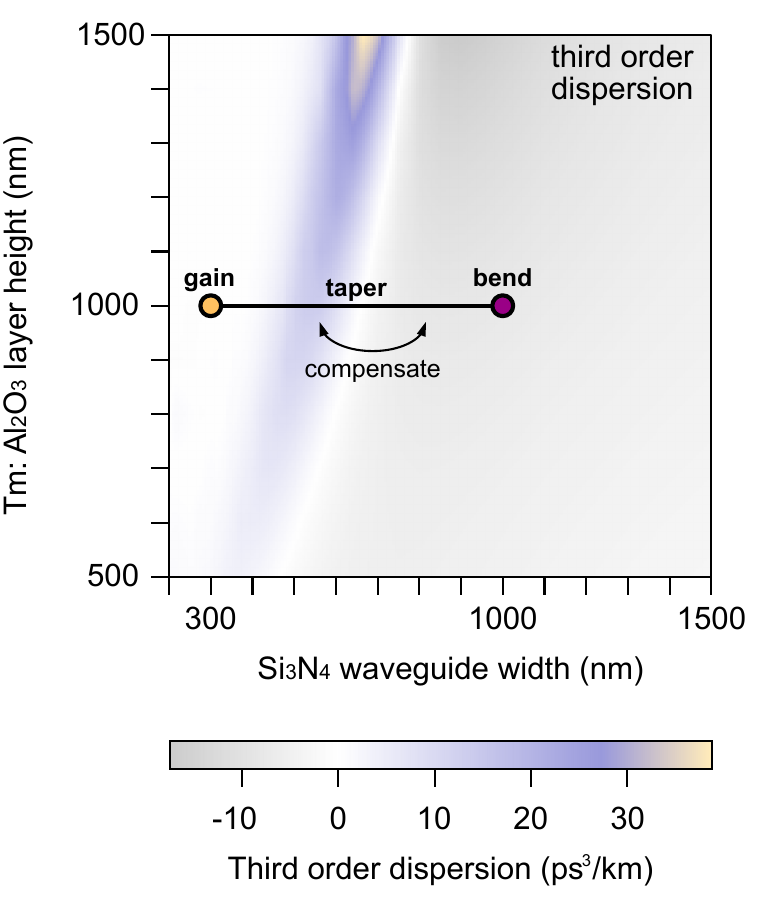}
  \makeatletter 
\renewcommand{\thefigure}{S\@arabic\c@figure}
\makeatother
  \caption{\textbf{Third order dispersion} in dependence of the silicon nitride waveguide width an the Al$_2$O$_3$ layer height. Along the taper, both negative and positive third order dispersion are present and can compensate each other. Specifically, the taper profile may be chosen to obtain a net-zero third order dispersion amplifier.
    }
\label{SI:fig:tdisp}
\end{figure}

For ultra-short pulses in the femtosecond regime, their large spectral bandwidth implies that third order dispersion effects become relevant. In our case, as evident in Fig.~3f in the main text, not perfectly compensated third order dispersion after compression leads to an asymmetric low intensity tail in the output pulse (and a `butterfly' shape in the FROG trace). Complementing main text Fig.~2b, Figure~\ref{SI:fig:tdisp} shows the third order dispersion in dependence of the silicon nitride waveguide width an the Al$_2$O$_3$ layer height. 
While the gain waveguides show close to zero third order dispersion, waveguide bends and tapers can exhibit considerable amounts of third order dispersion. Importantly, \textit{both} signs of third order dispersion occur along the taper, so that they can compensate each other. In our case, although peak dispersion values exceeding 10~ps$^3$/km occur along the taper and during the bend, their mean value is much smaller and the different contributions approximately compensate each other (mean third order dispersion along a taper ca. 1.6~ps$^3$/km; mean dispersion in a bend ca. 7~ps$^3$/km; per straight gain waveguide the amplifier comprises ca. $2\times3$~mm of taper and a bend length of ca. $1.5$~mm).

\section{Amplification without pre-chirping}
\begin{figure}[h!]
  \centering
\includegraphics[width=0.5\columnwidth]{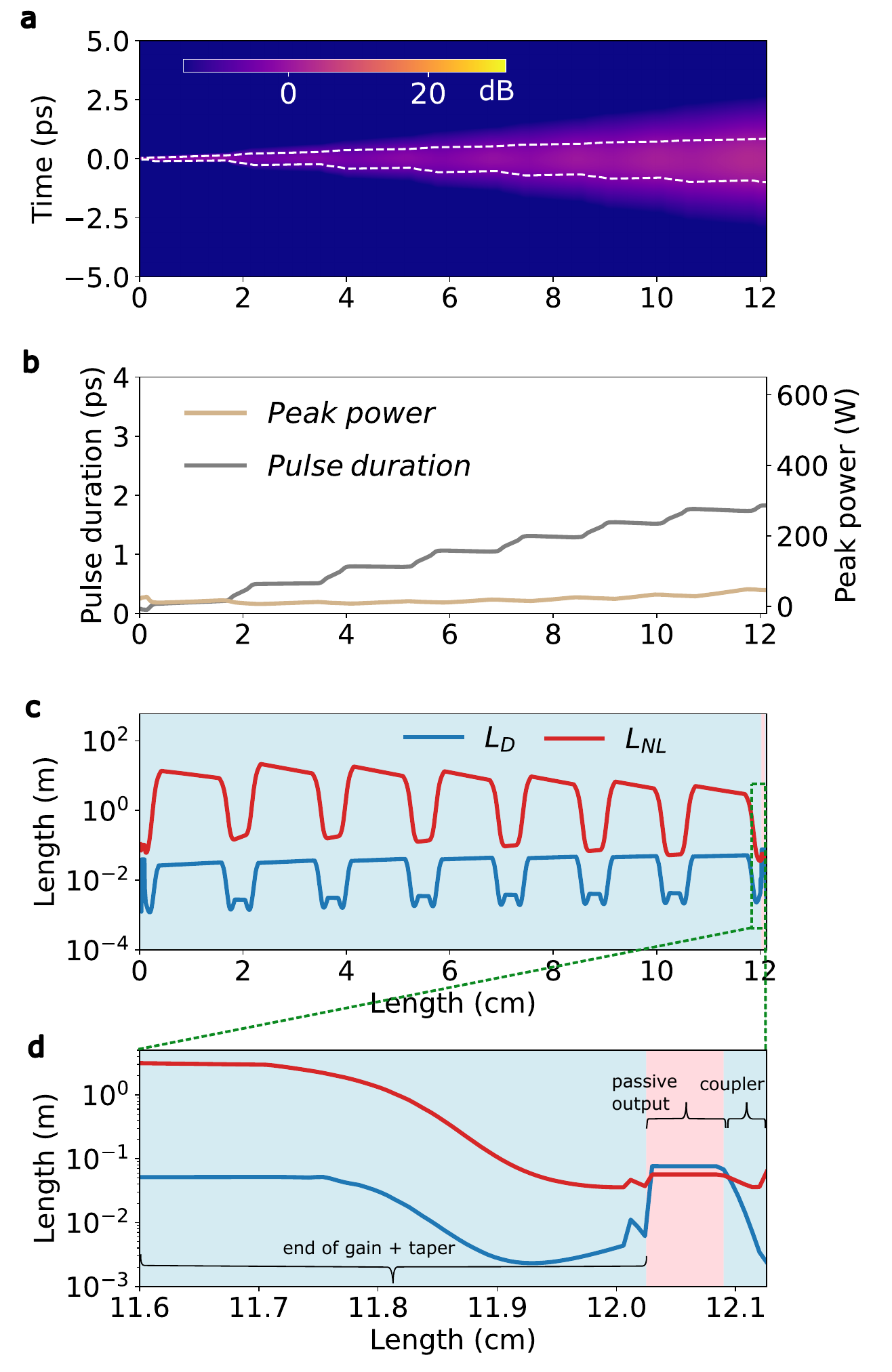}
  \makeatletter 
\renewcommand{\thefigure}{S\@arabic\c@figure}
\makeatother
  \caption{\textbf{Numerical simulation of signal amplification without pre-chirping.} \textbf{a}, Evolution of the temporal pulse power while propagating through the amplifier chip in a co-moving reference frame. One sequence of the alternating gain, bend and taper sections is indicated. The contours indicate the full-width-half-maximum (FWHM) pulse duration.
      \textbf{b}, Pulse duration and pulse peak power while propagating through the amplifier chip. \textbf{c}, Evolution of the pulse's dispersion length $L_\mathrm{D}$ and nonlinear length $L_\mathrm{NL}$ while propagating through the amplifier chip. The blue background color highlights where the propagation is dominated by linear optical effects ($L_\mathrm{D}<L_\mathrm{NL}$); nonlinear optical effects dominate only in short section within the last $1.5$~mm of the entire, $>$12~cm long propagation distance (red background color). \textbf{d}, shows a magnified portion of panel~c., where the gain section and taper to the output waveguide, the passive output waveguide, and the inverse taper for chip-output coupling are indicated.
    }
  \label{SI:fig:No prechirping}
\end{figure}

As mentioned in the main text, it is also possible to send the input pulses directly into the amplifier without pre-chirping. Due to the normal dispersion the pulses will spread in time before their gain in energy would cause substantial nonlinear effects. Figure~\ref{SI:fig:No prechirping} presents a numerical simulation of signal pulse amplification without pre-chirping the pulse, in the same representation as Fig.~4 in the main text). Figure~\ref{SI:fig:No prechirping}a reveals the temporal dynamics of amplification and pulse broadening due to normal dispersion of the amplifier. To obtain short pulses, dispersion compensation would now be required after amplification, at higher power levels. In contrast to pre-chirping at low power, this is more challenging due to the non-trivial power-dependent nonlinear compression dynamics.

\section{Critical power:  $L_\mathrm{NL}=L_\mathrm{D}$}
\begin{figure*}[h!]
  \centering
  \includegraphics[width=0.5\columnwidth]{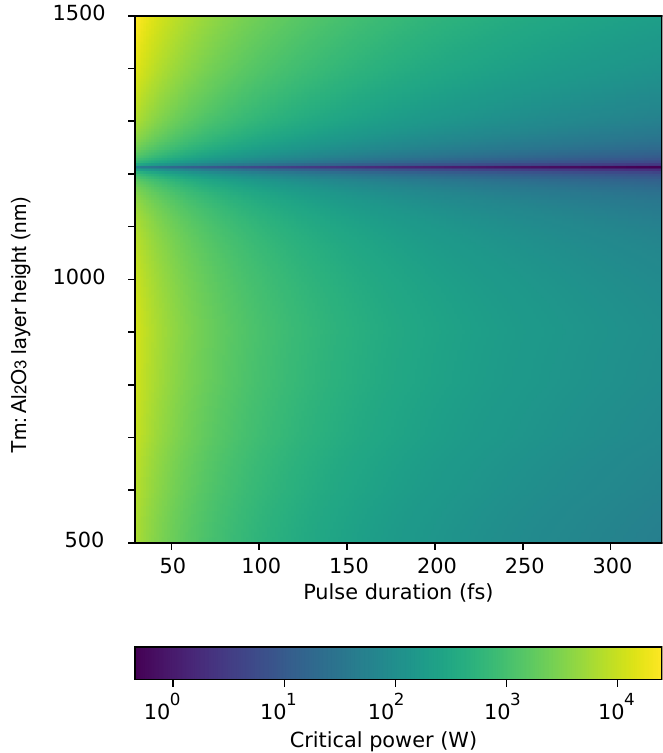}
    \makeatletter 
\renewcommand{\thefigure}{S\@arabic\c@figure}
  \caption{
      \textbf{Critical power} in dependence of the pulse duration and the Al$_2$O$_3$ layer height (assuming 300~nm wide silicon nitride waveguide). The achievable peak power scales quadratically with the inverse pulse duration and linearly with the (absolute) value of the GVD. For a layer height of $\sim1200$~nm, the GVD is close to zero and hence the critical power low. Higher critical power is achieved for shorter pulses and layer heights that exhibit larger GVD (cf. Fig.~2b in the main text).} 
  \label{fig:crit_power}
\end{figure*}

To provide an estimate of the maximally attainable peak power in a linear pulse propagation regime, we consider the critical power where nonlinear and dispersion length are equal $L_\mathrm{NL}=L_\mathrm{D}$. Figure~\ref{fig:crit_power} shows this critical power for different pulse duration (impacting $L_\mathrm{D}$) and gain layer height (impacting $L_\mathrm{D}$ and $L_\mathrm{NL}$), where we have assumed a fixed silicon nitride waveguide width of 300~nm. Figure~\ref{fig:crit_power} shows that a $\sim$100~fs pulse can be amplified and compressed to its time bandwidth limit with peak power levels of approximately 900~W, while remaining in the linear pulse propagation regime. For higher peak power levels nonlinearity will start contributing. Shorter pulses, which can in principle be supported by the Tm-doped gain medium could support higher peak power, in principle exceeding 10~kW, while remaining in the linear regime. To reach these power levels a careful management of higher order dispersion would be needed.

\end{document}